\documentclass{ifacconf}  
\usepackage{natbib}        
\usepackage[left=0.556in,right=0.556in,bottom=0.55in,top=1.45in]{geometry}

\usepackage{booktabs}
\usepackage{graphicx}

\usepackage{amssymb}
\usepackage{comment}
\usepackage{color,soul}
\usepackage{mwe}
\usepackage{svg}
\usepackage{mathtools}
\usepackage{balance}

\usepackage{cuted}

\usepackage{acronym}
\acrodef{OFO}{Online Feedback Optimization}

\usepackage[update,prepend]{epstopdf} 
\usepackage{fontawesome} 
\usepackage{circuitikz} 
\usepackage{tikz}
\usetikzlibrary{arrows.meta, decorations.markings}
\usepackage{cprotect} 
\usepackage[super]{nth}
\usepackage{subcaption}
\usepackage{algorithm,xpatch}
\usepackage{algpseudocode}
\usepackage{listings}
\usepackage{placeins}

\newtheorem{remark}{Remark}

\newcommand{\T}{^\mathsf{T}} 
\renewcommand{\d}{\mathrm{d}} 
\newcommand{\R}{\mathds{R}} 
\newcommand{\N}{\mathds{N}} 

\DeclareMathOperator*{\argmin}{arg\,min}

\usepackage{dsfont}

\usepackage{psfrag}
\usepackage{pstool}

\tikzstyle{block} = [draw, rectangle, 
    minimum height=3em, minimum width=5em]
\tikzstyle{sum} = [draw, circle, node distance=1cm]
\tikzstyle{input} = [coordinate]
\tikzstyle{output} = [coordinate]
\tikzstyle{tmp} = [coordinate]
\tikzstyle{pinstyle} = [pin edge={to-,thin,black}]

\xpatchcmd{\algorithmic}{\setcounter}{\algorithmicfont\setcounter}{}{}
\providecommand{\algorithmicfont}{}

\makeatletter
\let\save@mathaccent\mathaccent
\newcommand*\if@single[3]{%
  \setbox0\hbox{${\mathaccent"0362{#1}}^H$}%
  \setbox2\hbox{${\mathaccent"0362{\kern0pt#1}}^H$}%
  \ifdim\ht0=\ht2 #3\else #2\fi
  }
\newcommand*\rel@kern[1]{\kern#1\dimexpr\macc@kerna}
\newcommand*\wideaccent[2]{\@ifnextchar^{{\wide@accent{#1}{#2}{0}}}{\wide@accent{#1}{#2}{1}}}
\newcommand*\wide@accent[3]{\if@single{#2}{\wide@accent@{#1}{#2}{#3}{1}}{\wide@accent@{#1}{#2}{#3}{2}}}
\newcommand*\wide@accent@[4]{%
  \begingroup
  \def\mathaccent##1##2{%
    \let\mathaccent\save@mathaccent
    \if#42 \let\macc@nucleus\first@char \fi
    \setbox\z@\hbox{$\macc@style{\macc@nucleus}_{}$}%
    \setbox\tw@\hbox{$\macc@style{\macc@nucleus}{}_{}$}%
    \dimen@\wd\tw@
    \advance\dimen@-\wd\z@
    \divide\dimen@ 3
    \@tempdima\wd\tw@
    \advance\@tempdima-\scriptspace
    \divide\@tempdima 10
    \advance\dimen@-\@tempdima
    \ifdim\dimen@>\z@ \dimen@0pt\fi
    \rel@kern{0.6}\kern-\dimen@
    \if#41
      #1{\rel@kern{-0.6}\kern\dimen@\macc@nucleus\rel@kern{0.4}\kern\dimen@}%
      \advance\dimen@0.4\dimexpr\macc@kerna
      \let\final@kern#3%
      \ifdim\dimen@<\z@ \let\final@kern1\fi
      \if\final@kern1 \kern-\dimen@\fi
    \else
      #1{\rel@kern{-0.6}\kern\dimen@#2}%
    \fi
  }%
  \macc@depth\@ne
  \let\math@bgroup\@empty \let\math@egroup\macc@set@skewchar
  \mathsurround\z@ \frozen@everymath{\mathgroup\macc@group\relax}%
  \macc@set@skewchar\relax
  \let\mathaccentV\macc@nested@a
  \if#41
    \macc@nested@a\relax111{#2}%
  \else
    \def\gobble@till@marker##1\endmarker{}%
    \futurelet\first@char\gobble@till@marker#2\endmarker
    \ifcat\noexpand\first@char A\else
      \def\first@char{}%
    \fi
    \macc@nested@a\relax111{\first@char}%
  \fi
  \endgroup
}
\makeatother

\begin{document}
\begin{frontmatter}

\title{Adaptive Tuning of Online Feedback Optimization for Process Control Applications\thanksref{footnoteinfo}} 

\thanks[footnoteinfo]{Research supported by  Marie Curie Horizon Postdoctoral Fellowship project RELIC (M. Zagorowska and L. Imsland, grant no 101063948),  Delft Technology Fellowship (M. Zagorowska), and by the Wallenberg AI, Autonomous Systems and Software Program (WASP) funded by the Knut and Alice Wallenberg Foundation (G. Belgioioso).}

\author[First]{Marta Zagorowska} 
\author[Second]{Lukas Ortmann} 
\author[Third]{Giuseppe Belgioioso}
\author[Fourth]{Lars Imsland}

\address[First]{DCSC, TU Delft (e-mail: m.a.zagorowska@tudelft.nl).}
\address[Second]{Eastern Switzerland University of Applied Sciences (e-mail: lukas.ortmann@ost.ch)}
\address[Third]{Division of Decision and Control Systems, KTH Royal Institute of Technology, and Digital Futures (e-mail: giubel@kth.se)}
\address[Fourth]{Department of Engineering Cybernetics, Norwegian University of Science and Technology, (e-mail: lars.imsland@ntnu.no)}


\begin{abstract}
Online Feedback Optimization leverages properties of optimization algorithms to develop controllers for systems with limited model availability, which is often the case in process control. The interplay between the parameters of the chosen optimization algorithm, as well as lack of direct connection to the characteristics of the underlying process make their tuning challenging. We propose a method for adaptive tuning of Online Feedback Optimization controllers based on scaled projected gradient descent by using sensitivity of the desired objective to the parameters of the algorithm. The proposed adaptive tuning method limits the operator-tunable parameters to scalar values that represent how much the control inputs and the objective can change between iterations without requiring either additional information about the controlled system or repeated experiments. Numerical studies on a gas lift and a continuously-stirred tank reactor processes confirm that our adaptive scheme improves closed-loop performance of Online Feedback optimization compared to standard manual tuning methods.
\end{abstract}


\end{frontmatter}

\section{Introduction}
\label{sect:intro}
Online Feedback Opitmization (OFO) allows reaching the optimal operating points while satisfying constraints and relying on limited model information \citep{Optimization_Hauswirth2021}, which makes it a promising method for process control applications. The closed-loop performance of these controllers is determined by the parameters of the optimization algorithms, which are disconnected from the properties of the controlled system and usually need to be tuned through a series of experiments. To simplify tuning of these parameters, this paper proposes an adaptive method for choosing the parameters of Online Feedback Optimization removing the need for repeated experiments.

\subsubsection{Motivation}
Online Feedback Optimization based on scaled projected gradient descent has been proposed as a controller by \cite{Non_Haeberle2020}, with performance determined by two parameters: a scalar step size and a scaling matrix. Several methods have been proposed for adapting the scaling matrices in optimization algorithms, especially for unconstrained problems \cite[Chapter 5.6]{Introduction_Hazan2019}. However, OFO relies on real-time measurements and limited model information, which precludes the use of line-search strategies \citep{Non_Haeberle2020} or Hessian information \citep{scaled_Bonettini2008,Nonlinear_Bertsekas2016}. In practical uses of OFO, choosing the parameters requires multiple runs of the system \citep{Data_Gil2023,Tuning_Zagorowska2024}, or sufficient knowledge about the underlying system \citep{Tuning_Ortmann2024}.  \cite{Time_Bernstein2023} proposed an adaptive tuning method for the step size for steady-state problems with quadratic objective functions with linear constraints, while \cite{Feedback_Huang2025} also assumed linear dynamics of the underlying system. They used the derivatives with respect to problem variables as indicators whether the step should be increased or decreased. We build on these approaches and propose an online tuning method for nonlinear systems arising in process control applications.

\subsubsection{Contribution} 
In this work, we propose an adaptive method for choosing the scaling matrix and the step size to improve closed-loop performance and reduce the operator-tunable parameters to scalar values. We avoid additional experiments by approximating the derivatives with respect to parameters using known derivatives with respect to inputs and outputs that are used in standard OFO controllers. The scaling matrix is found by solving an auxiliary semi-definite optimization problem that preserves its properties, while the step size comes from a quadratic approximation of the objective inspired by \cite{Nonlinear_Bertsekas2016}.

This paper is structured as follows. Section \ref{sec:Background} introduces the background of OFO and the formulation used in this work. Section \ref{sec:AdaptiveTuning} introduces the adaptive tuning method, which is validated in numerical examples in Section \ref{sec:Numerical}. Section \ref{sec:conclusions} closes the paper with suggestions for future work.

\section{Background}
\label{sec:Background}
We consider the  steady-state optimization problem%
\begin{subequations} \label{eqn:ProblemStatement}
\begin{align}
\min_{u,y}& \quad \Phi(u,y)
    \label{eqn:CostFcn}\\
\text{s.t. }    & \quad y=h(u),\;Au\leq b,\;Cy\leq d,\label{eq:UBounds}
\end{align}
\end{subequations}
with $n_u$ inputs and $n_y$ outputs where $\Phi:\R^{n_u}\times\R^{n_y}\rightarrow\R$ is a continuously differentiable nonlinear cost, $h:\R^{n_u}\rightarrow\R^{n_y}$ is a continuously differentiable nonlinear input-output mapping, $A\in \R^{n_{c_1}\times n_u}$, $b\in\R^{n_{c_1}}$, $C\in\R^{n_{c_2}\times n_y}$, and $d\in\R^{n_{c_2}}$ are constant matrices, and $c_1$ and $c_2$ are the number of input and output constraints, respectively. 

\subsection{OFO with scaled projected gradient descent}
\label{sec:OFO}
Online Feedback Optimization with scaled projected gradient descent proposed by \cite{Non_Haeberle2020} solves \eqref{eqn:ProblemStatement} by iteratively finding control inputs as
\begin{align}
\label{eq:next_u_noG}
    u^{k+1} = u^k + \alpha^kw^k
\end{align} 
where $\alpha^k\in(0,\alpha_{\max}]$ is a tunable step size and \(w^k\) is the solution to the quadratic optimization problem:
\begin{subequations}
\label{eq:next_w_noG}
\begin{align}
    \min_{w\in\mathbb{R}^p} \quad &
    \left\| w + S^{k+1}H^\top(u^k)\nabla\Phi^\top(u^k,y^k)\right\|_{{(S^{k+1})}^{-1}}^2
    \label{eq:next_w_noG_obj}\\[3pt]
    \text{s.t.} \quad &
    A\left(u^k+\alpha_{\max}w\right)\leq b
    \label{eq:next_w_noG_con1}\\[3pt]
    &
    C \left(y^k+\nabla h(u^k)\alpha_{\max}w\right)\leq d
    \label{eq:YBound}\\[3pt]
    &
    S^k\in\mathbb{S}_+^{n_u},
    \label{eq:PositiveS}
\end{align}
\end{subequations}
where \(H(u)^\top = \left[\mathbb{I}_{n_u} \ \nabla h(u)^\top\right]\),  $S^k:\mathcal{U}\rightarrow\mathbb{S}_+^{n_u}$ is a continuous metric on $\mathcal{U}$, and $\|x\|_S:=\sqrt{x^\top S x}$. The set $\mathbb{S}_+^{n_u}$ denotes the set of symmetric positive definite matrices of size $n_u\times n_u$ and $\mathbb{I}_{n_u}$ is an identity matrix of size $n_u\times n_u$. Implementing \eqref{eq:next_u_noG} requires only the so-called sensitivity $\nabla h$ rather than the full input-output mapping $h$ from \eqref{eq:UBounds}, which is obtained as an on-line measurement $y^k$ in \eqref{eq:YBound}. 

The impact of the step $\alpha$ on closed-loop performance of the plant-algorithm interconnection was studied by \cite{Non_Haeberle2020}, while \cite{Tuning_Ortmann2024} have shown the role of the scaling matrix $S$. Typically, tuning these hyperparameters requires a series of experiments where the values are gradually increased until reaching the desired performance \citep{Data_Gil2023}\footnote{\cite{Data_Gil2023} considered $S=G^{-1}$, so the elements were decreased during tuning} and kept constant throughout online operation. However, repeated experiments are time-consuming and may be affected by changing operating conditions. 

\subsection{Sensitivity of Online Feedback Optimization}
\label{sec:Sensitivity}

Our adaptive tuning method leverages the sensitivity of the objective $\Phi$ in \eqref{eqn:CostFcn} to the scaling matrix $S$ and the step size studied by \cite{Sensitivity_Zagorowska}. From \eqref{eqn:CostFcn} and the chain rule we obtain at any time step $k\in \N_{+}$, $l\in\N_{\leq k}$:
\begin{equation}
\label{eq:PhiParamDeriv}
\begin{aligned}
     \frac{\partial\Phi^k}{\partial S^l} =&{} \left(\frac{\partial\Phi^k}{\partial u^{k}}+\frac{\partial\Phi^k}{\partial y^{k}}\cdot\frac{\partial y^k}{\partial u^{k}}\right)\cdot\frac{\partial u^k}{\partial S^l}.
    \end{aligned}
\end{equation}

To obtain $\frac{\partial u^k}{\partial S^l}$, we first get from \eqref{eq:next_u_noG}:
\begin{align}
    u^k=&{}u^0+\sum\limits_{l=0}^{k-1}\alpha^lw^l, \label{eq:nextvalue}
\end{align}
where $w^l\in\R^{n_u}$ is obtained from \eqref{eq:next_w_noG} and, as such, also depends on $S^0,\ldots, S^{l}$. By differentiating \eqref{eq:nextvalue} with respect to $S^k$, we get:
\begin{subequations}
\label{eq:duds}
\begin{align}
    \frac{\partial u^k}{\partial S^l}
    =&{} 
    \underbrace{\frac{\partial u^0}{\partial S^l}}_{=0}+\alpha^0\underbrace{\frac{\partial w^0}{\partial S^l}}_{=0}+\ldots
    +\alpha^l\frac{\partial w^l}{\partial S^l}+\ldots +\alpha^{k-1}\frac{\partial w^{k-1}}{\partial S^l}\nonumber\\
    =&{}\sum\limits_{m=l}^{k-1}\alpha^m\frac{\partial w^m}{\partial S^l}, \tag{\ref{eq:duds}}
\end{align}
\end{subequations}
because $u_0$ is independent of $S^l$ and $w^m$ is independent of $S^l$ for $m<l$. Stacking the columns of $S$ vertically into $p^l=\text{vec}(S^l)$ enables using the sensitivity formulas from \cite{Sensitivity_Zagorowska} to obtain $\frac{\partial w^m}{\partial S^l}$. 

\section{Adaptive tuning}
\label{sec:AdaptiveTuning}
The adaptive method in this work is inspired by iterative feedback tuning algorithms developed for PID controllers \citep{Iterative_Hjalmarsson1998,Iterative_Hjalmarsson2002}. These algorithms use sensitivity with respect to controller parameters to find optimal tuning values, but require a series of experiments to obtain derivatives of the objective with respect to individual controller parameters. 
We build on the assumption that OFO requires only the input-output sensitivity $\nabla h$ and exploit the sensitivity formulas from Section \ref{sec:Sensitivity} to adapt $S$ and $\alpha$ online.
\subsection{Scaling matrix adaptation}
Following \cite{Nonlinear_Bertsekas2016}, one of the goals in adapting the scaling matrix in gradient-based optimization is to ensure decrease of the objective between consecutive iterations:
\begin{equation}
\label{eq:WeWantThis}
    \Phi(u^{k+1},h(u^{k+1}))\leq \Phi(u^{k},h(u^{k})), \quad \forall k \in \mathbb{N}_+.
\end{equation}
If the input-output mapping $h$ is known, we can evaluate $\Phi(u^{k+1},h(u^{k+1}))$ to find $S$ that satisfies \eqref{eq:WeWantThis} using existing approaches based on Hessian information \citep[Section 2.3]{Nonlinear_Bertsekas2016}. However, since only the sensitivity $\nabla h$ and online measurements $y^k$ are available, we approach the problem using $u^{k+1}=u^k+\alpha_{\max}w^k$ and $y^{k+1}=y^k+\alpha_{\max}\nabla h(u^k)w^k$ to approximate \eqref{eq:WeWantThis}, yielding the approximated monotonic descent condition
\begin{equation}
\label{eq:WeWantThisFcnS}
    \Phi(u^k+w^k,y^k+\alpha_{\max}\nabla h(u^k)w^k)\leq \Phi(u^{k},y^{k}).
\end{equation}
We notice from \eqref{eq:next_w_noG},  that at time $k+1$, we note that $u^k$ and $y^k$ are known and $\Phi(u^{k+1},y^{k+1})$ is a function of $S^{k+1}$. Then \eqref{eq:WeWantThisFcnS} can be written as:
\begin{equation}
\label{eq:WeWantThisWithS}
    \Psi(S^{k+1})\leq \Psi(S^{k}).
\end{equation}
We now use the sensitivities from Section \ref{sec:Sensitivity} to approximate $\Psi(S^{k+1})$ around the current value $S^{k}$:
\begin{equation}
\label{eq:PsiApproximation}
    \Psi(S^{k+1})\approx\Psi(S^{k})+\Big\langle \frac{\partial\Phi^k}{\partial S^{k}}, S^{k+1}-S^{k}\Big\rangle_{\text{F}},
\end{equation}
where $\langle \cdot,\cdot\rangle_{\text{F}}$ is the Frobenius inner product \citep{horn2012matrix}. By substituting \eqref{eq:PsiApproximation} into \eqref{eq:WeWantThisWithS}, we obtain  
\begin{equation}
\label{eq:ConstraintExample}
    \Big\langle \frac{\partial\Phi^k}{\partial S^{k}},\, S^{k+1}-S^{k}\Big\rangle_{\text{F}}\leq 0,
\end{equation}
which characterizes the condition that \(S^{k+1}\) should satisfy to ensure monotonic decrease of the objective \eqref{eq:WeWantThis}.

\subsubsection{Optimization-based adaptation}

We note that \eqref{eq:ConstraintExample} is independent of $w^{k+1}$, thus we can find $S^{k+1}$ satisfying \eqref{eq:ConstraintExample} by solving an auxiliary optimization problem:
\begin{subequations}
\label{eq:FeasibilitySDP}
\begin{align}
    \min_{p,\,S} \quad &
    -p
    \label{eq:FeasibilitySDP_obj}\\[3pt]
    \text{s.t.} \quad &
    \Big\langle \frac{\partial\Phi^k}{\partial S^{k}},\, S-S^{k}\Big\rangle_{\text{F}}\leq -p,
    \label{eq:FrobeniusProductCstr}\\[3pt]
    & S\in\mathbb{S}_+^{n_u},    \label{eq:FeasibilitySDP_cstr1}\\
    & p\in[0,p_{\max}],
    \label{eq:FeasibilitySDP_cstr2}
\end{align}
\end{subequations}
where $p_{\max}>0$ limits how much the objective $\Phi$ is allowed to decrease, as indicated by \eqref{eq:PsiApproximation} and \eqref{eq:ConstraintExample}. To ensure that the scaling matrix stays strictly positive definite in \eqref{eq:FeasibilitySDP_cstr1}, for numerical purposes we introduce $\delta S=S^{k+1}-S^{k}$ and recast constraint \eqref{eq:PositiveS} equivalently as:
\begin{equation}
    \lambda_j(S^k+\delta S) \geq t, \quad j=1,\ldots, n_u,
    \label{eq:AllEigPositive}
\end{equation}
where $\lambda_j(X)$ denotes to the $j$-th eigenvalue of matrix $X$ and $t\geq t_{\min}>0$ is a constant ensuring strict positive definiteness. Rewriting \eqref{eq:AllEigPositive} as $S^k+\delta S - t\mathbb{I}\succcurlyeq 0$ allows expressing \eqref{eq:FeasibilitySDP_cstr1} in $\delta S$ that is symmetric but may be indefinite.

Furthermore, if \eqref{eq:FeasibilitySDP} admits a solution $S^{k+1}_*$, then there are infinitely many solutions, for example, those obtained by scaling $\kappa S^{k+1}_*$, with $\kappa >1$. Thus, we restrict $\delta S$ using the spectral norm $\|S^k+\delta S\|_{2}\leq t_{\max}$, where $t_{\max}>0$ is a design parameter. Then \eqref{eq:FeasibilitySDP} becomes:
\begin{subequations}
\label{eq:FeasibilitySDPFinalMinEig}
\begin{align}
    \min_{p,\,\delta S,\,t} \quad &
    -p - t
    \label{eq:Decreasingpandt}\\[3pt]
    \text{s.t.} \quad &
    \Big\langle \frac{\partial\Phi^k}{\partial S^{k}},\, \delta S\Big\rangle_{\text{F}} \leq -p,
    \label{eq:FinalDescent}\\[3pt]
    &
    S^k+\delta S - t\mathbb{I}\succcurlyeq 0,
    \label{eq:MinimalEigen}\\[3pt]
    &
    \|S^k+\delta S\|_{2}\leq t_{\max},
    \label{eq:limitingbyt}\\[3pt]
    &
    \delta S=\delta S^\top,\; p\in[0,p_{\max}],\; t\in[t_{\min},t_{\max}].
    \label{eq:FinalCond}
\end{align}
\end{subequations}
The constraint \eqref{eq:MinimalEigen} combined with the modified objective in \eqref{eq:Decreasingpandt} maximize the smallest eigenvalue of $S^k+\delta S$. The choice of $t_{\min}$ and $t_{\max}$, as well as the role of the spectral norm will be discussed in Section \ref{sec:AdaptationAlg}.

\subsubsection{Diagonal formulation}
Solving \eqref{eq:FeasibilitySDPFinalMinEig} allows finding a symmetric positive definite $S^{k+1}$ that improves the value of the objective by \eqref{eq:FinalDescent}. In practice, the scaling matrix is often restricted to diagonal, $S^k=\text{diag}(S^k_i)$ \citep{Tuning_Ortmann2024}. To further facilitate the implementation of OFO without solving \eqref{eq:FeasibilitySDPFinalMinEig}, we propose an adaptation method for this diagonal case.

The condition \eqref{eq:ConstraintExample} for the $i$-th element of $S^{k+1}$ is satisfied in two cases:
\begin{subequations}
\begin{align}
\frac{\partial\Phi^k}{\partial S_i^{k}}\leq 0,\;&{} S_i^{k+1}-S_i^{k}\geq 0 \label{eq:decreasingPhi},\\
\frac{\partial\Phi^k}{\partial S_i^{k}}\geq 0,\;&{} S_i^{k+1}-S_i^{k}\leq 0 \label{eq:increasingPhi}.
\end{align}
\end{subequations}
Thus, if $\Phi$ is increasing for $S_i^k$ in \eqref{eq:increasingPhi}, we can decrease the corresponding $S_i^{k+1}$ to avoid going too far from the current value. Conversely, if $\Phi$ is decreasing for $S_i^k$ as in \eqref{eq:decreasingPhi}, we can increase $S_i^{k+1}$ to reach the optimum more quickly. The proposed tuning rule is then:
\begin{subequations}
\label{eq:ProposedTuning}
\begin{align}
    \tilde{S}_i^{k+1}=&{}\begin{cases}
    (1+\beta_1)S_i^{k}\text{ if }\frac{\partial\Phi^k}{\partial S_i^{k}}< 0\\
        (1-\beta_2)S_i^{k}\text{ if }\frac{\partial\Phi^k}{\partial S_i^{k}}> 0,\\
        S_i^k\text{ otherwise}
    \end{cases}\\
    S_{i}^{k+1}=&{}\min\Big\lbrace t_{\max},\max \Big\lbrace t_{\min},\tilde{S}_i^{k+1}\Big\rbrace \Big\rbrace,
\end{align}
\end{subequations}
where $\beta_{1,2}\in(0,1)$ are design parameters. The tuning rule \eqref{eq:ProposedTuning} resembles a simplified two-way backtracking algorithm for scaled projected gradient descent from \cite{Backtracking_Truong2020,DeePCHuntData_Cummins2024}.

\subsection{Step size adaptation}
 To mitigate the impact of the potential inaccuracies of the approximation \eqref{eq:PsiApproximation}, we propose to adapt the step size $\alpha^k$ in \eqref{eq:next_u_noG} by approximating $\Phi(u^{k+1},y^{k+1})$ obtained for $S^{k+1}$ from \eqref{eq:FeasibilitySDPFinalMinEig} as a quadratic function of $\alpha$:
\begin{equation}
    \Phi(u^{k+1}(\alpha),y^{k+1}(\alpha)):=g(\alpha)=a_k\alpha^2+b_k\alpha +c_k,
\end{equation}
where parameters $a_k$, $b_k$, $c_k$ are obtained from a system of linear equations \cite[Section 1.2]{Nonlinear_Bertsekas2016}:
\begin{subequations}
\begin{align}
g(\Tilde{\alpha})=&{}\Phi(u^{k}+\Tilde{\alpha}w^k,y^k+\Tilde{\alpha}\nabla h(u^k))\label{eq:approximation},\\
\frac{\partial g}{\partial \alpha}(0)=&{}(H^\top(u_S^k)\nabla\Phi^\top(u_S^k,y_S^k))^\top w^k,\\
g(0)=&{}\Phi(u^k,y^k),
\end{align}
\end{subequations}
where $\Tilde{\alpha}\in[\alpha_{\min},\alpha_{\max}]$ is a chosen value. The approximated value in \eqref{eq:approximation} comes from approximating $y^{k+1}=y^k+\alpha^k\nabla h(u^k)w^k$. The step size $\alpha^{k+1}$ is then obtained by solving the following quadratic optimization problem:
\begin{equation}
\label{eq:StepMinimization}
    \alpha^{k+1}=\argmin_{\alpha\in[\alpha_{\min},\alpha_{\max}]} g(\alpha),
\end{equation}
with $\Tilde{\alpha}=\alpha^k$, and $\alpha_{\min},\alpha_{\max}\in \R_+$ describing the range of allowed step sizes. 

\subsection{Adaptation algorithm}
\label{sec:AdaptationAlg}

The proposed projected gradient OFO controller with adaptive tuning method is summarized in Algorithm \ref{alg:OFOiteration}. The adaptation is triggered if the optimum has not yet been reached with $\nabla \Phi(u^k,y^k)\neq 0$ and $w^k\neq 0$ (line 2). 

The algorithm requires choosing the scalar rparameters $p_{\max}$, $\alpha_{\min}$, $\alpha_{\max}$, $t_{\min}$, and $t_{\max}$. The value of $\alpha_{\max}$ indicates how good the linear approximation of $y^{k+1}$ in \eqref{eq:YBound} and in \eqref{eq:approximation} is. Ideally, $\alpha_{\max}$ should be chosen as a step size that would satisfy constraints in OFO with any $S^k=S$ such that $\|S\|_2\leq t_{\max}$. The value of the step size $\alpha_{\min}$ can be set to a small positive value, here $\alpha_{\min}=10^{-6}$.

The choice of $t_{\max}$ is related to the spectral norm properties in \eqref{eq:limitingbyt}. As $S^{k+1}\succ 0$, the constraint \eqref{eq:limitingbyt} limits the maximal eigenvalue of $S^{k+1}$, which in turn limits how much the matrix $S^{k+1}$ modifies both the direction and the length of the gradient $H^\top(u^k)\nabla\Phi^\top(u^k,y^k)$ in the projection \eqref{eq:next_w_noG} \citep{Linear_Strang2000}. Thus, $t_{\max}$ should be chosen to allow overcoming small $H^\top\nabla\Phi^\top$. Conversely, $t_{\min}$ relates to the smallest eigenvalue of $S^{k+1}$. If $t_{\min}\geq 1$, then solving \eqref{eq:FeasibilitySDPFinalMinEig} is only allowed to lengthen $H^\top\nabla\Phi^\top$, while $0\leq t_{\min}<1$ allows shortening it, effectively reducing the step size. In this work $t_{\min}=10^{-6}$ while $t_{\max}$ is chosen depending on the case study.

{\linespread{1}
\begin{algorithm}[!tbp]
 \caption{OFO iteration with adaptive tuning\label{alg:OFOiteration}}
 \begin{algorithmic}[1]
    \Require $p_{\max}$, $t_{\min}$, $t_{\max}$, $\alpha_{\min}$, $\alpha_{\max}$, $S^{0}$
    \If{$(H^\top(u^k)\nabla\Phi^\top(u^k,y^k))^\top w^k< 0$} 
        \State Solve \eqref{eq:FeasibilitySDPFinalMinEig}, set $S^{k}=S^{k-1}+\delta S$
    \EndIf
    \State Solve \eqref{eq:next_w_noG}, get $w^{k}$
    \State Solve \eqref{eq:StepMinimization}, get $\alpha^{k}$
    \State Compute $u^{k+1}$ from \eqref{eq:next_u_noG}, apply, and measure $y^{k+1}$
    \State Compute $\frac{\partial\Phi^k}{\partial S^{k}}$ from \eqref{eq:PhiParamDeriv}
    \item \Return $\frac{\partial\Phi^k}{\partial S^{k}}$, $u^{k+1}$, $y^{k+1}$, $w^{k}$
    \end{algorithmic}
\end{algorithm}
}

\section{Numerical examples}
\label{sec:Numerical}

\subsection{Impact of adaptation}
\label{sec:ToySetup}

\subsubsection{Impact of scaling matrix}
\begin{figure}[!tbp]
         \centering
         \includegraphics[width=0.4\textwidth]{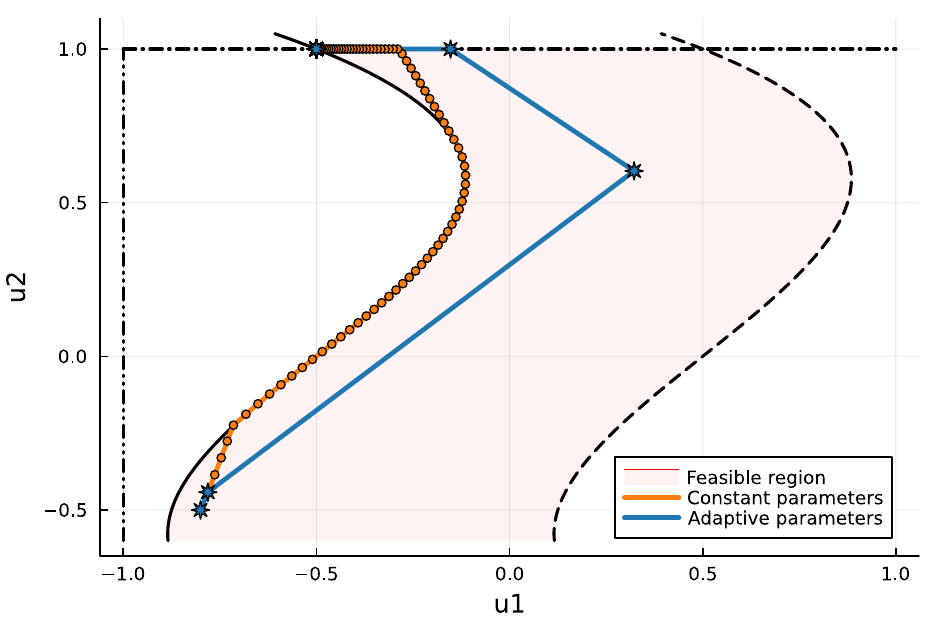}
         \caption{Adaptation accelerates OFO for the example from \cite{Non_Haeberle2020} while still respecting non-convex feasible region (pink)}
         \label{fig:BetterThanVHpaper}
\end{figure}

The first example is taken from \cite{Non_Haeberle2020} with the mapping $y=h(u)=u_2^3+u_1-u_2+0.5$ where the objective is to minimise $\Phi(u,y)=1.5u_1^2+u_2^2-u_2^3+u_1u_2-3u_2+1.5+y$ over $u_1,u_2\in[-1,1]$, $y\in[0,1]$. The expected behaviour is to reach $(u_1^*,u_2^*)=(-0.5,1)$ with $\Phi^*=-1.625$ starting from $(u^0_1,u^0_2)=(-0.8,-0.5)$ with $\Phi^0=4.81$ within $t_f=100$ iterations. For comparison of the approach from \cite{Non_Haeberle2020} obtained for fixed $S=\mathbb{I}_2$ and $\alpha=0.01$, we set $p_{\max}=1$, $S^0=\mathbb{I}_2$, $\alpha^0=\alpha_{\max}= 0.01$, $t_{\max}=1000$ in Algorithm \ref{alg:OFOiteration}. Figure \ref{fig:BetterThanVHpaper} confirms that the adaptation accelerates OFO, allowing reaching the optimum in five (orange) instead of 82 (blue) iterations while respecting the non-convex feasible region (pink).

\subsubsection{Impact of step size}
\begin{figure}[!tbp]
     \centering
          \begin{subfigure}[b]{0.38\textwidth}
         \centering
         \includegraphics[width=\textwidth]{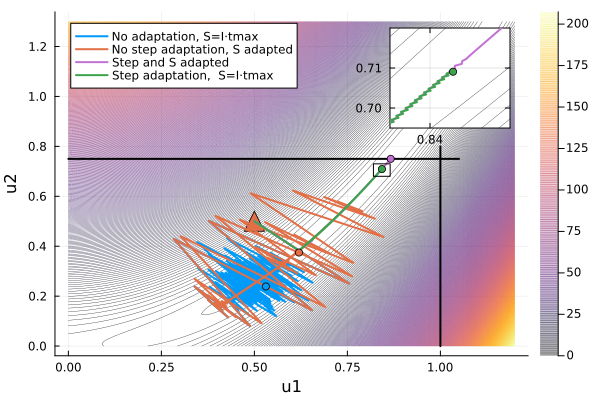}
         \caption{Contour plot and trajectories}
         \label{fig:StepSizeAdaptationSth}
     \end{subfigure}
     \begin{subfigure}[b]{0.38\textwidth}
         \centering
         \includegraphics[width=\textwidth]{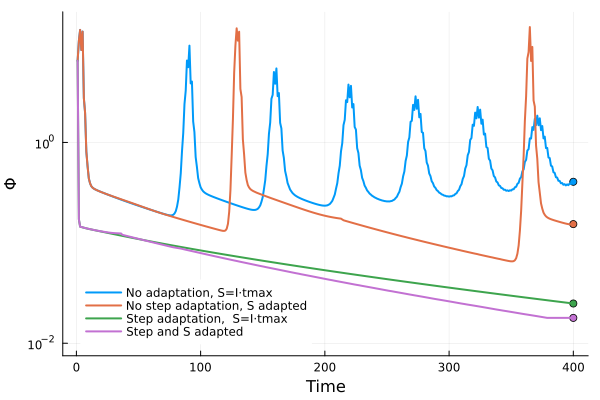}
         \caption{The objective function}
         \label{fig:StepSizeAdaptation}
     \end{subfigure}
  \begin{subfigure}[b]{0.38\textwidth}
         \centering
         \includegraphics[width=\textwidth]{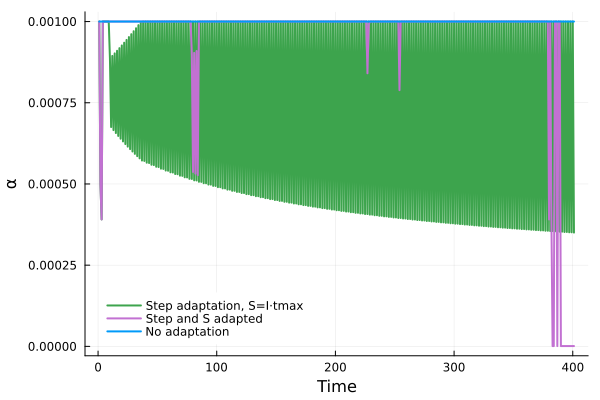}
         \caption{Step size adaptation}
         \label{fig:StepSizeActual}
     \end{subfigure}
        \caption{Impact of step size adaptation for the Rosenbrock function}
        \label{fig:StepAdaptation}
\end{figure}

We show the impact of the step size adaptation \eqref{eq:StepMinimization} on a Rosenbrock function, adapted to the OFO framework with $y=[y_1(u),y_2(u)]^\top=[10(u_2-u_1^2),1-u_1]^\top$ and $\Phi(u,y)=y_1^2+y_2(1-u_1)$. We also set $u_1\in[-1,1]$, $u_2\in[-1,0.75]$ and $y_i\in [-5,5]$, $i=1,2$ to allow reaching the optimum $u^*=[0.86,0.75]$ (circle) within 400 timesteps from $u_0=[0.5,0.5]^\top$ (triangle in Fig. \ref{fig:StepSizeAdaptationSth}). The initial condition and the constraints were chosen to include the valley-like part of the Rosenbrock function. We set $t_{\max}=5$, $p_{\max}=1$, and $\alpha_{\max}=0.001$.

We show the impact of the step size adaptation from \eqref{eq:StepMinimization} in four cases (Fig. \ref{fig:StepSizeAdaptation}): No adaptation and $S^k=\text{const}=\text{diag}(t_{\max})$ (blue), step size adaptation and $S^k=\text{const}=\text{diag}(t_{\max})$ (green), no step size adaptation and $S^k$ adapted (orange), both step size and $S^k$ adapted (purple). We see that using \eqref{eq:FeasibilitySDPFinalMinEig} requires the step adaptation \eqref{eq:StepMinimization}, otherwise descent is not guaranteed (orange). At the same time, doing only step adaptation from \eqref{eq:StepMinimization} while keeping $S^k=\text{const}=\text{diag}(t_{\max})$ improves the solution because the modifications to the gradient due to $S^k$ are compensated for by taking smaller steps (green). Combining the scaling matrix and the step adaptation results in fast convergence without oscillations (purple).

Figure \ref{fig:StepSizeActual} shows how the step size was adapted across the timesteps. Both cases with the step adaptation were able to prevent the oscillatory behaviour. However, a closer inspection shows small scale oscillations (green in the inset in the top right corner in Fig. \ref{fig:StepSizeAdaptationSth}). The step size adaptation \eqref{eq:StepMinimization} tries to compensate for the fixed scaling matrix by modifying how far the inputs move in a given direction. The adaptation of the step size has to compensate for large changes to the gradient induced by a fixed $S=t_{\max}\mathbb{I}$, because the eigenvalues $\lambda_i(S)=t_{\max}$ increase the effective step size at every timestep as indicated by \eqref{eq:next_u_noG} (Fig. \ref{fig:StepSizeActual}). A larger step size combined with the valley of the Rosenbrock function introduces the small oscillations visible in the inset in Fig. \ref{fig:StepSizeAdaptationSth}. In contrast, adapting the scaling matrix prevents oscillations as the scaled gradient gets aligned towards the descent direction (purple).

\subsection{Gas lift optimization with diagonal scaling}
\label{sec:NoCoupling}

\begin{figure}[!tbp]
     \centering
         \centering
         \includegraphics[width=0.38\textwidth]{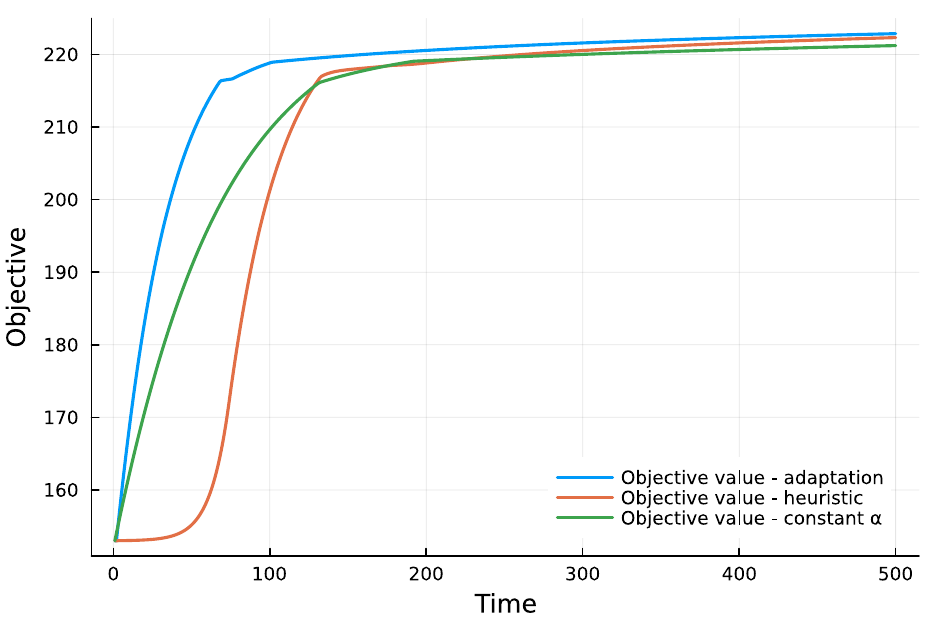}
         \caption{Comparison of convergence for the gas lift case study}
         \label{fig:GasLiftComparison_all1}
\end{figure}

The gas lift example comes from \cite{Data_Andersen2023} and was first adapted to the Online Feedback Optimization framework by \cite{Sensitivity_Zagorowska}. The objective is to maximise the cumulative output of two floating oil platforms $\max_{u,y} y_1(u)+y_2(u)
$, ${u}=[u_i]_{i=1,\ldots,5}$,${y}=[y_i]_{i=1,2}$ connected to two $y_1({u})=f_1({u})+f_2({u})$ and three oil wells $y_2({u})= f_3({u})+f_4({u})+f_5({u})$ where $f_i({u})$ describes the characteristics of the $i$-th well as a function of the amount of natural gas, $u_i$, injected into the well to facilitate oil extraction. The inputs and the outputs are bounded $u_i\in[\underline{u}_i,\overline{u}_i]_{i=1,\ldots,5}$, $y_i\in[\underline{y}_i,\overline{y}_i]_{i=1,2}$, and the overall amount of available gas is limited through a coupling constraint, $\sum_{i=0}^5 u_i\leq 26000$  Sm$^{3}$day$^{-1}$. The initial value was set $u_0=[2500.0,7000.0,4500.0,4500.0,4500.0]$ Sm$^3$day$^{-1}$. The case study is put in \eqref{eqn:ProblemStatement} as $\Phi({u},{y})=-y_1({u})-y_2({u})$, $h({u})=\begin{bmatrix}f_1({u})+f_2({u}) & f_3({u})+f_4({u})+f_5({u})\end{bmatrix}^{\T}$, and the details were provided by \cite{Sensitivity_Zagorowska}. We restrict $S$ to diagonal to emulate the structure of the Hessian of $\Phi$, and set $t_{\max}=1000$ and $\alpha_{\max}=1$. The value of $p_{\max}$ is set to 10 to reflect limitations on the oil production from timestep $k$ to $k+1$. 

\subsubsection{Comparison with fixed step size}
The comparison with results obtained for $\alpha^k=\text{const}=500$ from \cite{Sensitivity_Zagorowska} is shown in Fig. \ref{fig:GasLiftComparison_all1}. Both Algorithm \ref{alg:OFOiteration} and the heuristic \eqref{eq:ProposedTuning} reached the optimum more quickly than if constant $\alpha$ was used, even after the constraints became active (timestep 68 for the optimization-base approach, 130 for the heuristic approach).

\subsubsection{Adaptation analysis}

\begin{figure}[!tbp]
     \centering
     \begin{subfigure}[b]{0.38\textwidth}
         \centering
         \includegraphics[width=\textwidth]{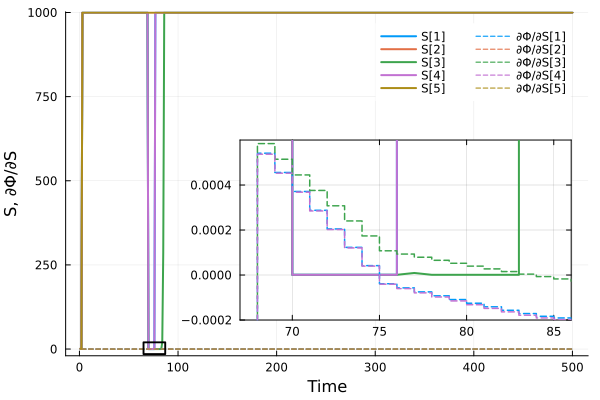}
         \caption{Using Algorithm \ref{alg:OFOiteration} }
         \label{fig:SMatrixAdaptationOpt}
     \end{subfigure}
          \begin{subfigure}[b]{0.38\textwidth}
         \centering
         \includegraphics[width=\textwidth]{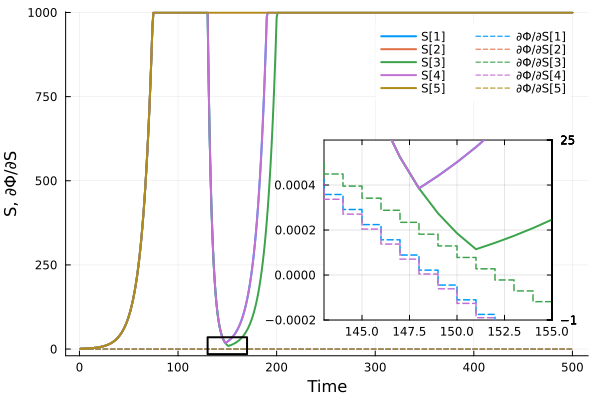}
         \caption{Using the heuristic from \eqref{eq:ProposedTuning}}
         \label{fig:SMatrixAdaptationHeuristic}
     \end{subfigure}
        \caption{Adaptation of the matrix $S$ for the gas lift case study}
        \label{fig:Adaptation}
\end{figure}

Figure \ref{fig:Adaptation} shows how the adaptation works when solving \eqref{eq:FeasibilitySDPFinalMinEig} (Fig. \ref{fig:SMatrixAdaptationOpt}) and using the heuristic \eqref{eq:ProposedTuning} with $\beta_1=0.1$ and $\beta_2=0.2$ (Fig. \ref{fig:SMatrixAdaptationHeuristic}). Both tuning approaches follow a similar pattern of increasing the elements of $S^{k+1}$ when the problem is unconstrained (time below 50 in Fig. \ref{fig:GasLiftComparison_all1}). When the coupling constraint  becomes active (timestep 68 for the optimization-base approach, 130 for the heuristic approach), the derivative \eqref{eq:PhiParamDeriv} becomes greater than zero (dashed lines in Fig. \ref{fig:Adaptation}), prompting the decrease of the elements of $S^{k+1}$ to satisfy \eqref{eq:FinalDescent} (solid lines in Fig.  \ref{fig:SMatrixAdaptationOpt}) and \eqref{eq:increasingPhi} (solid lines in Fig.  \ref{fig:SMatrixAdaptationHeuristic}). When the derivative becomes negative again, both approaches increase the elements of $S^{k+1}$, thus increasing the step size and allowing reaching the optimum faster. 

The main difference between Algorithm \ref{alg:OFOiteration} and the heuristic is in the rate of adaptation of the elements of $S^k$ (time from zero to 50 in Figs. \ref{fig:GasLiftComparison_all1} and \ref{fig:Adaptation}). In Algorithm \ref{alg:OFOiteration}, the adaptation of $S^k$ from $k$ to $k+1$ is restricted only by $t_{\max}$ through \eqref{eq:limitingbyt}. For instance, the fourth element of $S^k$ in Fig. \ref{fig:SMatrixAdaptationOpt} goes from $t_{\max}=1000$ to $t=10^{-6}$ in timestep 70 and then from $t$ to $t_{\max}$ at timestep 76. The rate of change in the heuristic approach is limited by the choice of $\beta_1$ and $\beta_2$ in \eqref{eq:ProposedTuning}. As a result, at the beginning the heuristic approach needs 70 iterations to bring the elements of $S^k$ to $t_{\max}$. As $\beta_2>\beta_1$, the heuristic algorithm needs 50 iterations to decrease the fourth element from $t_{\max}$ to $t_{\min}$ due to positive derivatives (around iteration 150 in Fig \ref{fig:SMatrixAdaptationHeuristic}).


\subsection{Comparison with manual tuning for dynamic systems}
In this case study we show that the proposed algorithm allows avoiding repeated experiments when applied to a dynamic system. As a case study, we use a continuously stirred tank reactor (CSTR) with nonlinear van der Vusse reaction adapted from \cite{RobustnessConsiderationsPID_Alfaro2012}:
\begin{subequations}
\label{eq:DynamicCSTR}
\begin{align}
\frac{\d c_A(t)}{\d t}=&{}\frac{F(t)}{V}(c_{Ai}(t)-c_A(t))-k_1c_A(t)-k_3c_A^2(t),\\
\frac{\d c_B(t)}{\d t}=&{}-\frac{F(t)}{V}c_B(t)+k_1c_A(t)-k_2c_B(t),
\end{align}
\end{subequations}
where $c_A$, $c_B$ are concentrations of reactants $A$ and $B$, respectively, $V$ is the constant volume of the reactor, $k_i$, $i=1,2,3$ are reaction rates. The two concentrations $c_A$ and $c_B$ are the outputs, $y=[c_A,c_B]^{\T}$, while the inlet flow rate $F$ and the concentration of reactant $A$ in the inlet stream $c_{Ai}$ are inputs, $u=[F,c_{Ai}]^{\T}$. The initial conditions together with the values of constant parameters are in Table \ref{tbl:CSTRParameters}.

\begin{table*}[!tbp]
\centering
\caption{Parameters for the CSTR case study, adapted from \cite{RobustnessConsiderationsPID_Alfaro2012}}
\label{tbl:CSTRParameters}
\begin{tabular}{l|lllllll}
Name & $V$     & $k_1$          & $k_2$          & $k_3$                     & $c_A^{\min}$, $c_B^{\min}$, $F^{\min}$, $c_{Ai}^{\min}$ & $c_A^{\max}$ & $c_B^{\max}$ \\
Value & 700 l & 5/6 min$^{-1}$ & 5/3 min$^{-1}$ & 1/6 l mol$^{-1}$ min$^{-1}$ & 0                                               & 10  mol l$^{-1}$       & 5    mol l$^{-1}$     \\ \midrule
Name  &$c_{Ai}^{\max}$ & $F^{\max}$ & $F_0$    & $c_{Aio}$ & $c_{A0}$ & $c_{B0}$&$\Delta T$ \\ 
Value &15 mol l$^{-1}$            & 634 l min$^{-1}$     & 350.15 l min$^{-1}$& 10.15 mol l$^{-1}$   & 2.82 mol l$^{-1}$  & 1.08 mol l$^{-1}$ & 1 min  \\ 
\end{tabular}
\end{table*}

The objective is to adjust $F$ and $c_{Ai}$ so that the concentration $c_B$ reaches desired setpoint $r$: $\Phi(u,y)=(c_B-r)^2$. The input-output sensitivity $\nabla h$ is obtained implicitly from steady state conditions of \eqref{eq:DynamicCSTR}:
\begin{subequations}
\begin{align}
0=&{}\frac{F}{V}(c_{Ai}-c_A)-k_1c_A-k_3c_A^2,\\
0=&{}-\frac{F}{V}c_B+k_1c_A-k_2c_B.
\end{align}
\end{subequations}
The outputs, $c_A$ and $c_B$, and the inputs, $F$ and $c_{Ai}$, are bounded (Table \ref{tbl:CSTRParameters}).  The upper bounds $F^{\max}$ and $c^{\max}_{Ai}$ on the inputs $F$ and $c_{Ai}$ indicate that the two inputs can have different orders of magnitude, which require adjusting individual elements in the scaling matrix for tuning \citep{Tuning_Ortmann2024}. Finally, while OFO can be applied without timescale separation \citep{Stability_Bianchi2024}, application of OFO to continuous systems requires setting the control interval $\Delta T$ between two consecutive iterations $k$ and $k+1$ \citep{Tuning_Zagorowska2024}, here $\Delta T=1$ min.

\begin{figure*}[!tbp]
     \centering
     \begin{subfigure}[b]{0.31\textwidth}
         \centering
         \includegraphics[width=\textwidth]{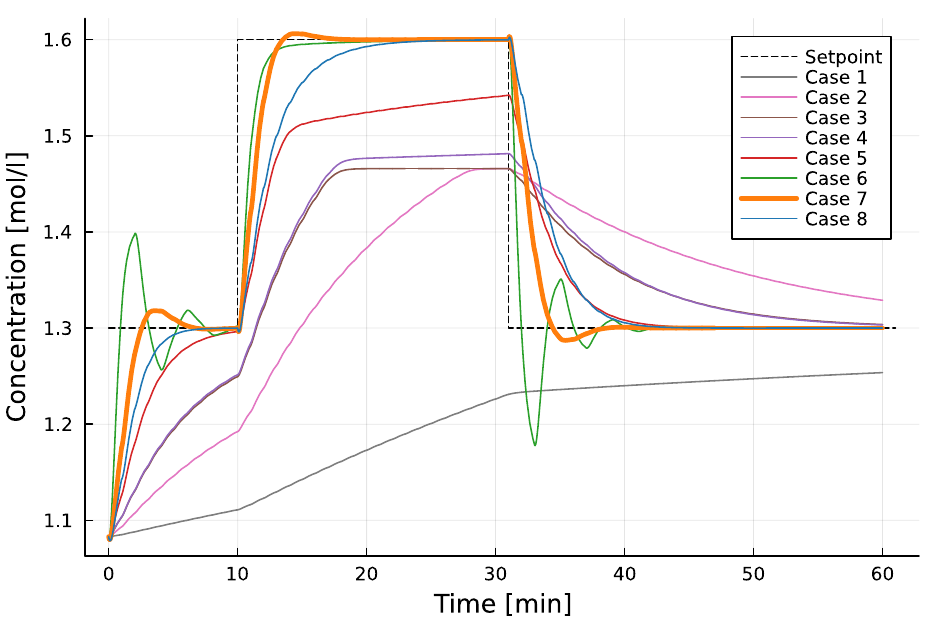}
         \caption{Manual tuning trajectories with the chosen case in bold orange}
         \label{fig:TuningTrajectoriesAll}
     \end{subfigure}
     \begin{subfigure}[b]{0.31\textwidth}
         \centering
         \includegraphics[width=\textwidth]{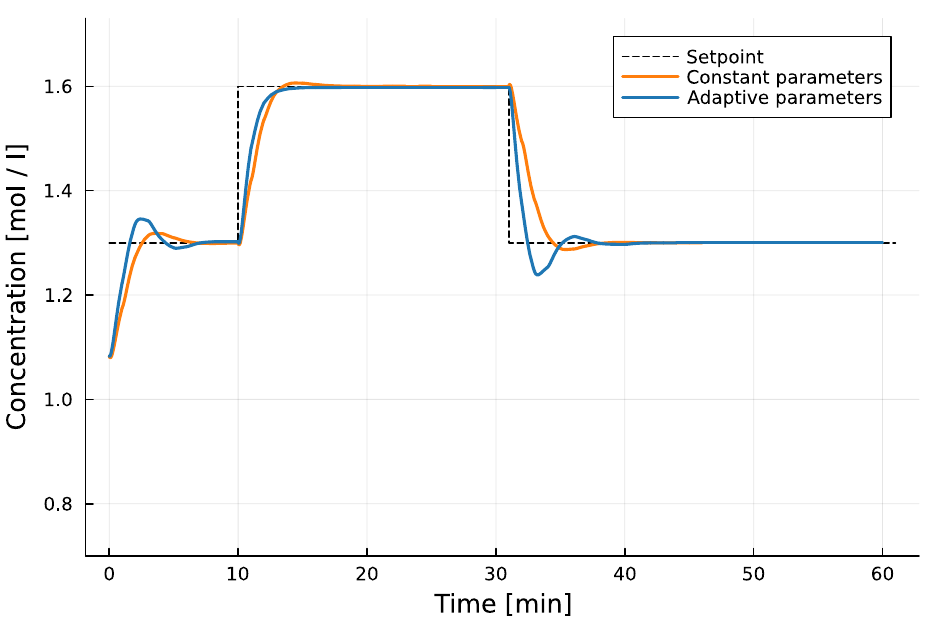}
         \caption{Comparison on the tuning setpoint - trajectories }
         \label{fig:TuningTrajectories}
     \end{subfigure}
          \begin{subfigure}[b]{0.31\textwidth}
         \centering
         \includegraphics[width=\textwidth]{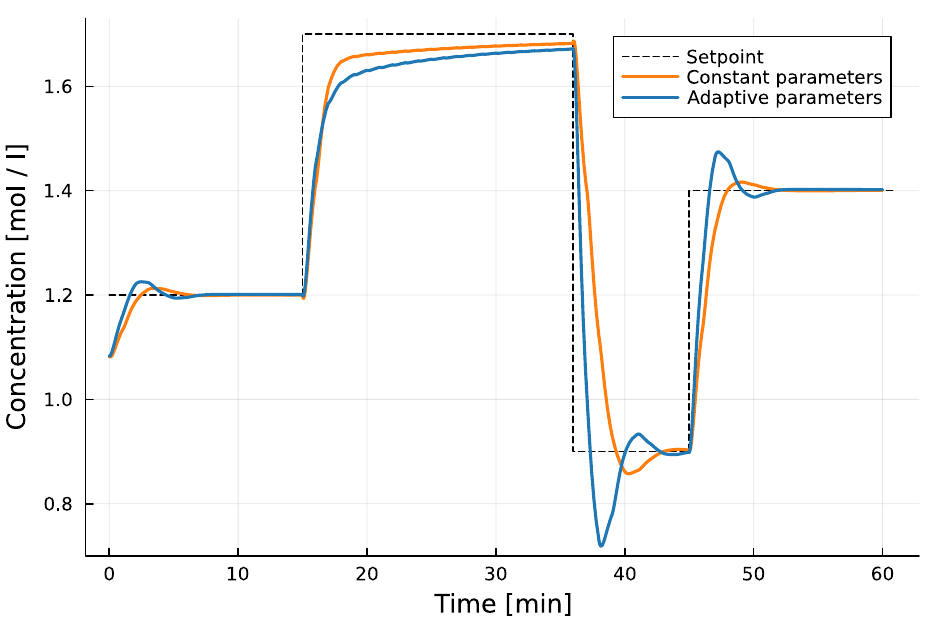}
         \caption{Comparison on the validation setpoint - trajectories}
         \label{fig:ValidationTrajectories}
     \end{subfigure}
     \begin{subfigure}[b]{0.31\textwidth}
         \centering
         \includegraphics[width=\textwidth]{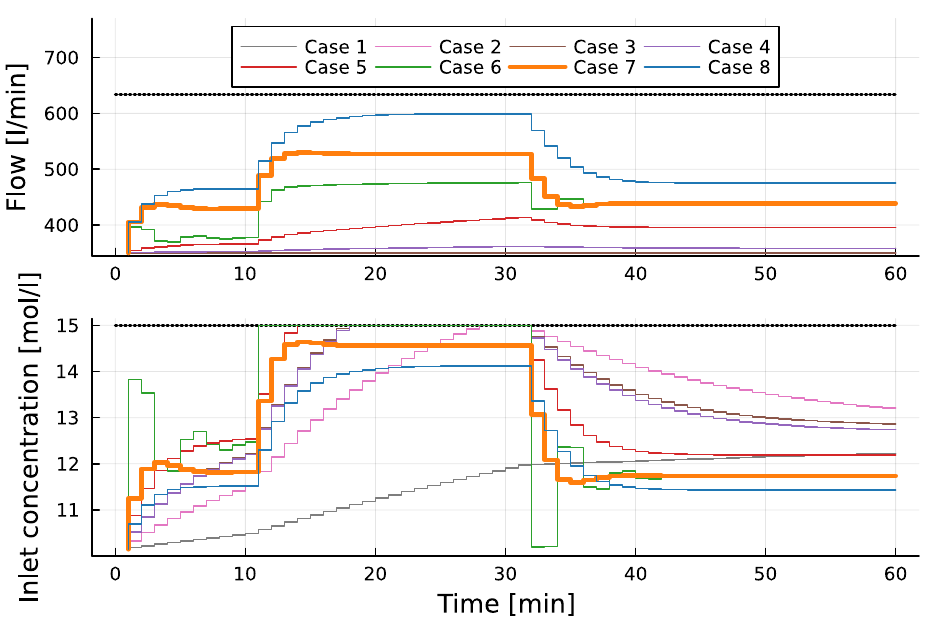}
         \caption{Manual tuning inputs with the chosen case in bold orange}
         \label{fig:TuningInputsAll}
     \end{subfigure}
          \begin{subfigure}[b]{0.31\textwidth}
         \centering
         \includegraphics[width=\textwidth]{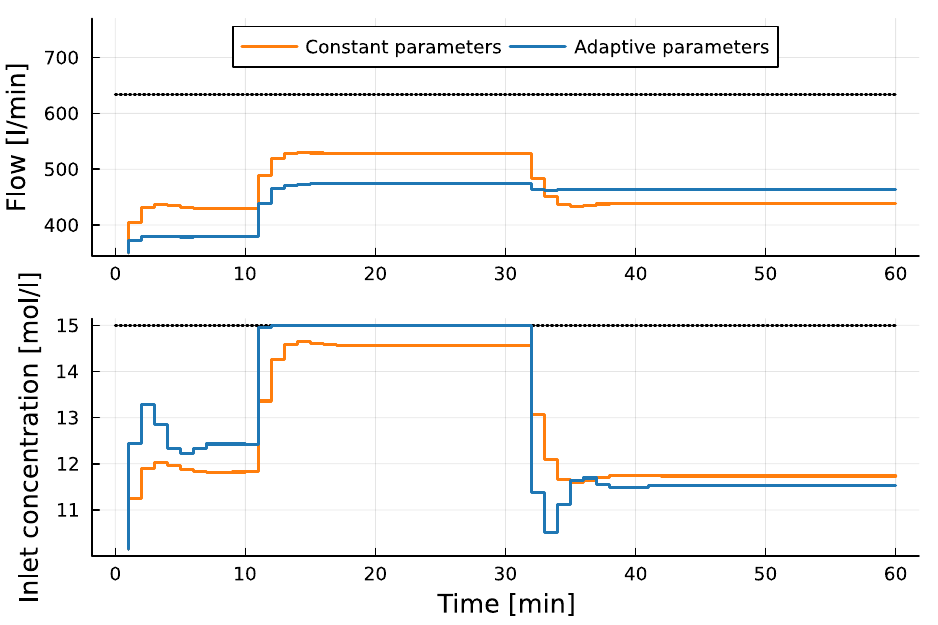}
         \caption{Comparison on the tuning setpoint - inputs}
         \label{fig:TuningInputs}
     \end{subfigure}
          \begin{subfigure}[b]{0.31\textwidth}
         \centering
         \includegraphics[width=\textwidth]{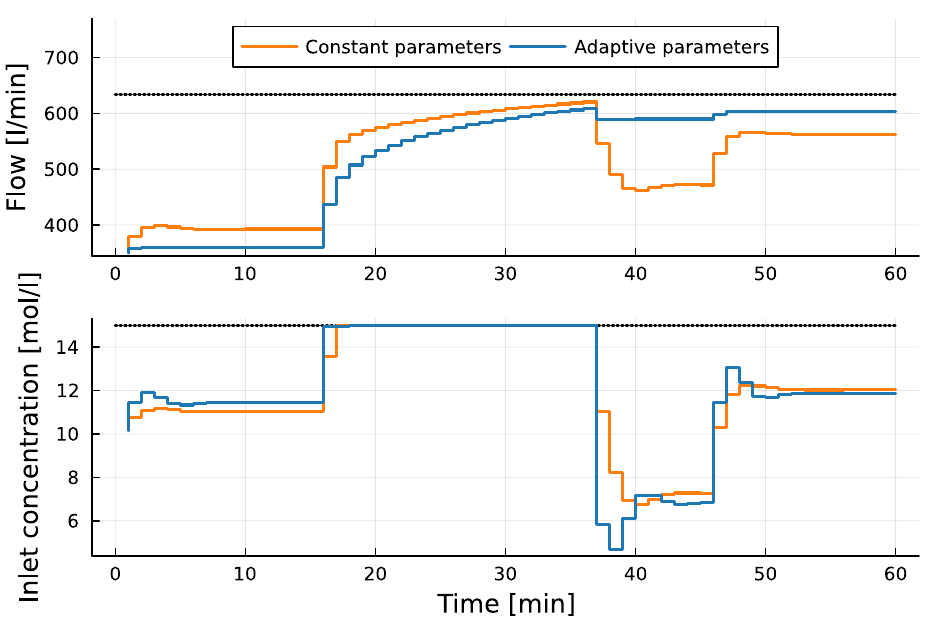}
         \caption{Comparison on the validation setpoint - inputs}
         \label{fig:ValidationInputs}
     \end{subfigure}
        \caption{CSTR case study results: multiple experiments during manual tuning (Figs. \ref{fig:TuningTrajectoriesAll} and \ref{fig:TuningInputsAll}), comparison of the parameters chosen manually and using Algorithm \ref{alg:OFOiteration} on the set-point used for tuning (Figs. \ref{fig:TuningTrajectories} and \ref{fig:TuningInputs}), comparison of the parameters chosen manually and using Algorithm \ref{alg:OFOiteration} on a different validation set-point (Figs. \ref{fig:ValidationTrajectories} and \ref{fig:ValidationInputs})}
        \label{fig:CSTRAdaptation}
\end{figure*}

We compare Algorithm \ref{alg:OFOiteration} with manual tuning proposed by \cite{Data_Gil2023}. The results of the manual tuning are in Fig. \ref{fig:TuningTrajectoriesAll} (trajectories $c_B$) and \ref{fig:TuningInputsAll} (inputs $F$ and $c_{Ai}$). The values of parameters of the eight experimental cases needed for manual tuning are in Table \ref{tbl:CSTRTuningManual}. As expected, small values (Cases 1, 2, 3) lead to sluggish responses, and insufficient performance is confirmed by the error values:
\begin{equation}
\label{eq:CSTRError}
    \epsilon = \frac{1}{N}\sum\limits_{i=1}^N (y_i-r_i)^2,
\end{equation}
where $N=60$ denotes the time horizon. The sluggish response is explained by the inputs in Fig. \ref{fig:TuningTrajectories}, where small values of parameters lead to insufficient use of the flow $F$ (top), while quickly saturating the concentration $c_{Ai}$ (bottom). Conversely, increasing the values decreases the error, at the expense of oscillations (Case 6). In particular, increasing the element of the matrix $S$ that corresponds to the flow and decreasing the element corresponding to concentration allowed using the first input without saturating the second one (Cases 7 and 8). The analysis of the error combined with a visual inspection of the responses in Fig. \ref{fig:TuningTrajectoriesAll} led to setting Case 7 as the chosen values (bold orange).

The performance of OFO with the chosen parameters was then compared to OFO with the adaptive Algorithm \ref{alg:OFOiteration} on the same tuning trajectory (Figs. \ref{fig:TuningTrajectories} and \ref{fig:TuningInputs}) and on a different validation trajectory (Figs. \ref{fig:ValidationTrajectories} and \ref{fig:ValidationInputs}). The parameters of the adaptive algorithm were chosen as $\alpha_{\max}=120$ and $t_{\max}=1000$. The adaptation algorithm (blue) allowed reaching 25\% better tracking accuracy on the tuning trajectory (Table \ref{tbl:CSTRErrorValues}) without repeated experiments. The error from \eqref{eq:CSTRError} obtained for the validation  trajectory in Fig. \ref{fig:ValidationTrajectories} indicates that the adaptation algorithm outperforms manual tuning by up to 20\% (Table \ref{tbl:CSTRErrorValues}). The impact of adaptation is primarily visible when large step changes occur at times 35 min and 45 min. By adjusting the matrix and the step size, the algorithm is able to follow the reference more closely than manually tuned OFO.

\begin{table*}[!tbp]
\centering
\caption{Manual tuning over eight experiments, with the chosen case in \textbf{bold}}
\label{tbl:CSTRTuningManual}
\begin{tabular}{@{}lllllllll@{}}
\toprule
         & Case 1 & Case 2 & Case 3 & Case 4                                     & Case 5       & Case 6        & \textbf{Case 7}           & Case 8            \\ \midrule
$\alpha$ & 1      & 5      & 5      & 10                                         & 20           & 100           & 120              & 120               \\
$S$      & $\mathbb{I}_{2}$      & $\mathbb{I}_{2}$      & $0.5\mathbb{I}_{2}$   & $\begin{bmatrix}100& 0\\ 0& 1\end{bmatrix}$ & $\begin{bmatrix}500& 0\\ 0 &1\end{bmatrix}$ & $\begin{bmatrix}1000 & 0\\ 0 &1\end{bmatrix}$ & $\begin{bmatrix}1000& 0\\ 0 &0.25\end{bmatrix}$ & $\begin{bmatrix}1000& 0\\ 0& 0.125\end{bmatrix}$ \\
Error& 0.083 & 0.033 & 0.019&  0.019 & 0.012 & 0.007 & \textbf{0.011} & 0.012\\
\bottomrule
\end{tabular}
\end{table*}


\begin{table}[!tbp]
\centering
\caption{Error values obtained using \eqref{eq:CSTRError} for tuning and validation trajectories in the CSTR case study (relative percentage between adaptive and manual in parentheses)}
\label{tbl:CSTRErrorValues}
\begin{tabular}{@{}l|ll@{}}
           & Manual               & Adaptive            \\
           \hline
Tuning     & 0.011 (100\%) & 0.008 (74.5\%) \\
Validation & 0.047 (100\%)   & 0.037 (79\%)
\end{tabular}
\end{table}

\section{Conclusions and future works}
\label{sec:conclusions}
A controller based on Online Feedback Optimization with scaled projected gradient descent allows reaching the optimum of an objective function while satisfying constraints without needing exact models of the controlled systems. The closed-loop performance of the controllers is defined by the parameters of optimization algorithms, the step size and the scaling matrix, often requiring multiple experiments for manual tuning. In this work, we replace the manual tuning of Online Feedback Optimization based on projected gradient descent with a one-time choice of scalar parameters that describe allowed changes in the inputs and in the objective function for the controlled system. Instead of using experiments to gain information about the influence of the step size and scaling matrix, the proposed method uses sensitivity of the objective to find values providing descent towards the optimum. The sensitivities are then used to solve an auxiliary semi-definite optimization problem to find the scaling matrix online. Numerical examples confirm that the online adaptation allows OFO to reach the optimum more quickly compared to manual tuning, and without additional experiments. 

The proposed algorithm relies on approximations of the objective and the constraints, and ongoing work focuses on providing theoretical guarantees of convergence and constraint satisfaction for Online Feedback Optimization with the proposed tuning. Moreover, the tuning method proposed in this work relies on sensitivity of the objective to parameters, so robustifying against real-life noisy measurements will further facilitate practical implementation.

\section*{DECLARATION OF GENERATIVE AI AND AI-ASSISTED TECHNOLOGIES IN THE WRITING PROCESS}
During the preparation of this work the author(s) did not use any Generative AI or AI-assisted technologies.

\balance

\appendix
\section{Adaptation of individual elements of the matrix}
\begin{remark}
    \label{rem:Counterexample}
The requirement of a diagonal matrix $S$ is necessary to ensure that $S^{k+1}\in\mathbb{S}_+^{n_u}$. As a counter-example, let us consider $S^{k}=\begin{bmatrix}
    0.11&-0.1\\-0.1&0.1
\end{bmatrix}\in \mathbb{S}_+^{n_u}$ with eigenvalues $\lambda_{1,2}(S^k)=0.0048,0.205$. We adapt $S^k$ according to \eqref{eq:ProposedTuning} with $\beta_{1,2}=0.1$ and obtain $S^{k+1}=\begin{bmatrix}
    0.11\cdot 0.9&-0.1\\-0.1&0.1
\end{bmatrix}\notin \mathbb{S}_+^{n_u}$ with eigenvalues $\lambda_{1,2}(S^{k+1})=-0.0005,0.1995$.

\end{remark}

\begin{remark}
\label{rem:IFT}
If we set $\beta_1=-\beta_2=\frac{\partial\Phi^k}{\partial S_i^{k}}\cdot\left(S_i^{k}\right)^{-1}$ in \eqref{eq:ProposedTuning}, we obtain an analogue of the method from \cite[Eq. (30)]{Iterative_Hjalmarsson2002} for Online Feedback Optimization.
\end{remark}

The counterexample from Remark \ref{rem:Counterexample} shows that the approach from \cite{Iterative_Hjalmarsson2002} summarised in Remark \ref{rem:IFT} cannot be used directly if controller parameters are matrices.

\section{Numerical setup}
The numerical examples were run on Windows 10 Enterprise in Julia 1.10.5 \citep{bezanson2017julia} using Clarabel v0.9.0 \citep{Clarabel_2024}, ForwardDiff v0.10.36 \citep{RevelsLubinPapamarkou2016}, JuMP v1.23.2 \citep{Lubin2023}, and ImplicitDifferentiation v0.6.0 \citep{ImplicitDifferentiation.jl}.

\end{document}